\documentclass[twocolumn,showpacs,preprintnumbers,amsmath,amssymb,floatfix]{revtex4}
\usepackage{graphicx}

\begin{document}
\newcommand{\bit}{\begin{itemize}}
\newcommand{\eit}{\end{itemize}}
\newcommand{\bc}{\begin{center}}
\newcommand{\ec}{\end{center}}
\newcommand{\be}{\begin{equation}}
\newcommand{\ee}{\end{equation}}
\newcommand{\beqn}{\begin{eqnarray}}
\newcommand{\eeqn}{\end{eqnarray}}
\newcommand{\ba}{\begin{array}}
\newcommand{\ea}{\end{array}}
\newcommand{\ra}{\rightarrow}
\newcommand{\lra}{\longrightarrow}
\newcommand{\bt}{\begin{tabular}}
\newcommand{\et}{\end{tabular}}

\title {Critical properties of loop percolation models 
with optimization constraints}

\author{
Frank O. Pfeiffer
and Heiko Rieger
}

\address{
Theoretische Physik, Universit\"at des Saarlandes,
       66041 Saarbr\"ucken, Germany
}
\date{\today}

\begin{abstract}
We study loop percolation models in two and in three space dimensions,
in which configurations of occupied bonds are forced to form closed
loop. We show that the uncorrelated occupation of elementary plaquettes
of the square and the simple cubic lattice by elementary loops leads
to a percolation transition that is in the same universality class as
the conventional bond percolation. In contrast to this an optimization
constraint for the loop configurations, which then have to minimize a
particular generic energy function, leads to a percolation transition
that constitutes a new universality class, for which we report the
critical exponents. Implication for the physics of 
solid-on-solid and vortex glass models are discussed.\\
\end{abstract}

\pacs{64.60.Ak, 64.60.Cn, 64.60.Fr}   


\maketitle

\section{Introduction}

The percolation of loops (or closed strings) appears naturally in the
context of liquid helium \cite{Wil89,Scha01}, early universe
\cite{Wil99,SH97} and high-temperature superconductors
\cite{NS98,PR02}, where loops represent world-lines, cosmic strings and
vortex loops, respectively. In analogy to the characteristic size of a
cluster in conventional site or bond percolation \cite{Sta85BH96},
the typical diameter $\xi$ of the loops diverges when approaching a
critical point, the loop percolation transition. This transition shows
power-law behavior at the critical point, which is described by a set
of critical exponents that constitute a universality class.

In this paper we study these percolation transitions numerically.
We show that the loop percolation (LP) model, in which
each elementary plaquette of a square lattice (in $2d$) or on a simple
cubic lattice (in $3d$) is occupied with a probability $p$ with an
elementary loop, is in the same universality as the conventional
$d$-dimensional bond percolation. More importantly we also show that in
contrast to the LP model an optimization constraint for the loop
configurations, which then have to minimize a particular generic
energy function, leads to a percolation transition that constitutes a
new universality class, for which we report the critical exponents.

These {\it loop Hamiltonian} (LH) models are relevant for the ground state
properties of disordered solid-on-solid models \cite{RB97,PR00,ZKNM98}
and vortex glasses \cite{KR98}. One particular example of was recently
studied by us in the context of a $3d$ vortex glass model for amorphous
high-$T_c$ superconductors in the strong screening limit\cite{PR02},
which shows an unconventional percolation transition of vortex loops
in the ground state as a function of the disorder strength $\sigma$.

The paper is organized as follows: In section \ref{models}, we
introduce the models and the definition of {\it loops} (clusters). In
section \ref{results} we locate the percolation transition in each
model and calculate the critical exponents using finite-size scaling
(FSS).  Section \ref{summary} concludes the paper with a summary and
a discussion.

\section{Models}
\label{models}

\begin{figure}[h]
\includegraphics[width=\columnwidth]{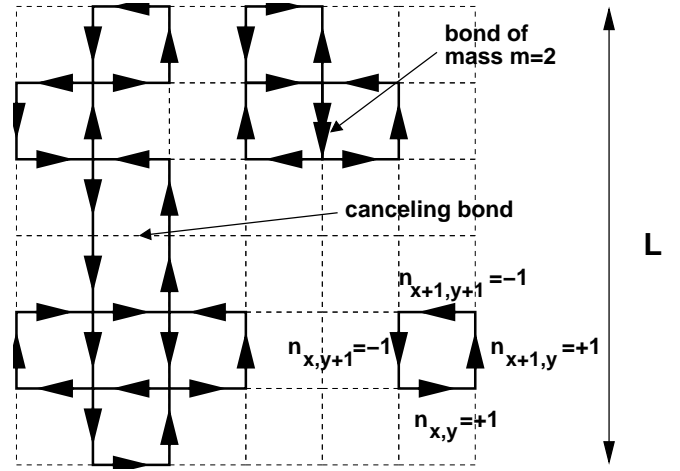}
\caption{
Configuration ${\bf n}$ of the loop percolation model on a $2d$ square lattice
with system size $L=7$.
}
\label{loopconf}
\end{figure}

Consider a $d$-dimensional hyper-cubic lattice - i.e. a square lattice
in $2d$ or a simple cubic lattice in $3d$ - of linear size $L$ ($L=7$ in
the $2d$ example in Fig.\ref{loopconf}) with free boundary conditions (f.b.c.).
In the loop percolation (LP) model the elementary plaquettes of the lattice
are occupied with elementary loops with a probability $p_1$. 
(There are $(L-1)^2$ elementary plaquettes in $2d$ and $3L(L-1)^2$
elementary plaquettes in $3d$.)
An elementary loop consists of the four bonds belonging to an
elementary plaquette plus a randomly chosen direction: either clockwise or
counterclockwise both with probability $1/2$ (see Fig. \ref{loopconf}).
When two adjacent plaquettes are occupied by elementary loops of same
orientation we cancel the occupation of the common bond as indicated
in Fig. \ref{loopconf}.

%
\begin{figure*}
\includegraphics[width=15cm]{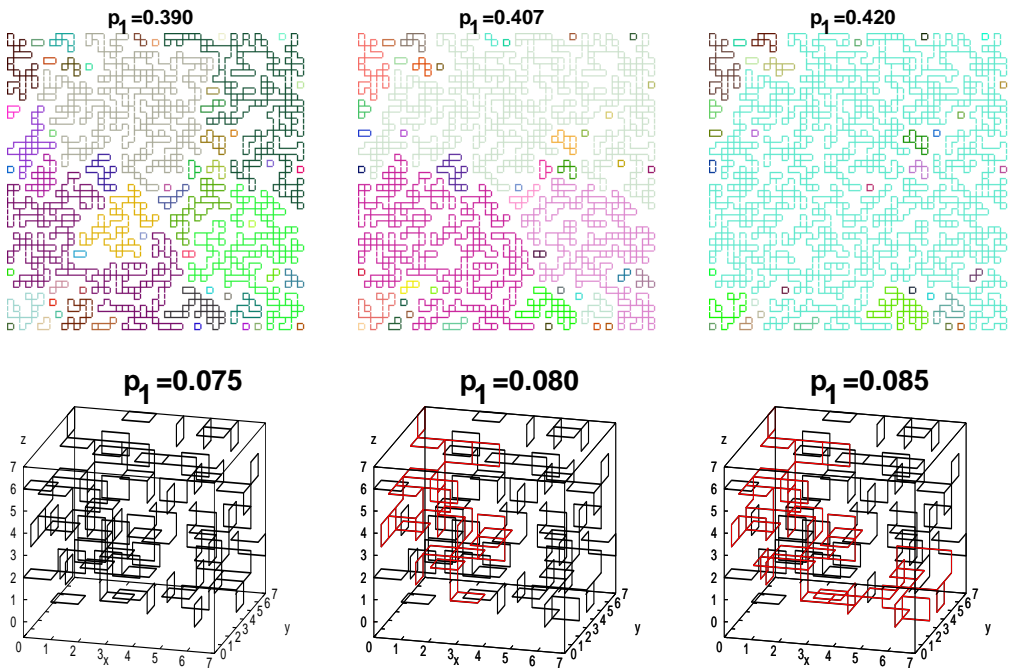}
\caption{
Typical loop configurations of the LP model around the
critical point $p_{1c} \approx 0.41$ in $2d$ (top) for $L=50$ and
$p_{1c} \approx 0.08$ in $3d$ (bottom) for $L=8$.  In $2d$ the
different loops are marked by different grey scales (colors), whereas
in $3d$ all loops are black except for the percolating loop, which is
marked by light grey (red).  }
\label{conf_3d3}
\end{figure*}

We can identify the resulting (directed) bond configuration of the LP
model with a flow ${\bf n}= \{ n_1, n_2, ..., n_M\}$, where $n_i$ is
an integer and $M$ is the number of bonds in the lattice.  We say that
$n_i=0$ if bond $i$ is not occupied, $n_i=\pm 1$ if it is singly
occupied in the positive (negative) direction (positive and negative
are defined by the introduction of an appropriate coordinate system),
$n_i=\pm 2$ if it is doubly occupied etc.  Thus, an elementary loop
(e.g. in the $xy$ plane) can be represented by $n_{x,y} = n_{x+1,y} =
-n_{x+1,y+1} = -n_{x,y+1} = 1$ if oriented counterclockwise, as shown
in Fig. \ref{loopconf}. The complete flow ${\bf n}$ then can simply
be thought of as the sum of all elementary loops.  Obviously, in this
sum the flow variables on the common bonds of adjacent elementary
loops cancel arithmetically.  Moreover, the construction of this flow
via addition of elementary loops implies that on each site of the
lattice the number of ingoing arrows balances the number of outgoing
arrows (see Fig. \ref{loopconf}): one says that this flow is
divergence free
\be
\nabla \cdot n_i =0.
\label{divfree}
\ee
%
(The lattice-divergence operator is defined on each lattice site and sums
simply all $2 \cdot d$ flow variables of the bonds connected to it).

In analogy to conventional bond percolation we are now going to define
the clusters of a configuration of the LP model: Two occupied bonds
belong to the same cluster if they have one site in common.  Thus, all
bonds of the cluster can be connected via a directed path along
occupied bonds belonging to this cluster, which is analogous to
conventional bond percolation - up to the attribute {\it directed}.
This is actually a slightly non-trivial observation - see
Fig. \ref{loopconf} to exemplify this statement - because an occupied
directed bond cannot be traversed in the opposite direction, but is a
direct consequence of the fact that a cluster is also a sum of
elementary loops.

\begin{figure*}
\includegraphics[width=14cm]{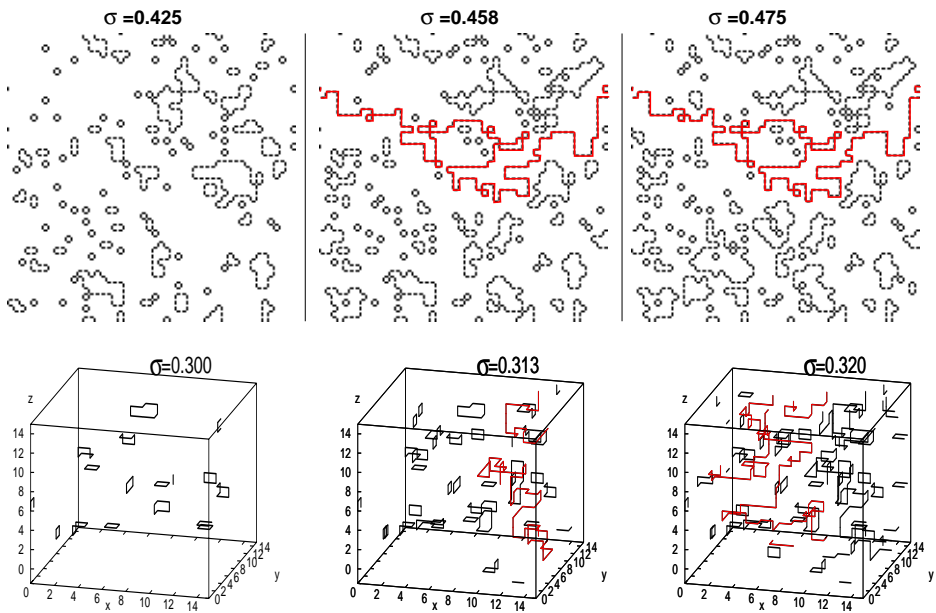}
\caption{
Typical loop configurations of the loop Hamiltonian (LH) model in the
ground state ($T=0$) around the critical point $\sigma_c \approx 0.46$
in $2d$ (top) for $L=50$ and $\sigma_c \approx 0.31$ in $3d$ (bottom) for $L=16$.
In $2d$ and $3d$ all loops are black except for the percolating loop, which is
marked by light grey (red).
}
\label{conf_hm}
\end{figure*}

The mass $m$ of a cluster is the number of occupied bonds, where a
bond $i$ with a flow $n_i$ counts $|n_i|$ times, i.e. has mass
$m_i=|n_i|$.  A percolating cluster is a cluster spanning the entire
system (in at least one of the $d$ directions).  This implies that the
cluster contains a directed path along its occupied bonds from one
side of the lattice to the opposite. In the following we refer to the
cluster just defined as a loop.

The representation of the configuration of directed occupied bonds as
a flow ${\bf n}$ is now used to define the LP model {\it with} an
optimization constraint: In contrast to the stochastic {\it
uncorrelated} occupation of elementary loops in the LP model above, we
now consider a different type of occupation of loops, which results
from the minimization of an energy function for a loop configuration
${\bf n}$
\be
H=H\{{\bf n}\}=\sum_{i=1}^M f_i(n_i)\;,
\ee
where the sum is over all bonds $i$ on a $d$-dimensional hyper-cubic
$L^d$ lattice with periodic boundary conditions (p.b.c).  Herewith we
assume the energy function to be composed of solely of local terms
$f_i(n)$ with $f_i(n)\ge0$ and convex (i.e.\ $f''(n)\ge0$) for all
bonds $i$ and flow (or occupation) values $n$. Such energy functions
are relevant in the context of disordered solid-on-solid models (in
$2d$) \cite{RB97,PR00,ZKNM98} and vortex glasses (in $3d$) \cite{KR98}
since they determine their ground states. One has to keep in mind,
that the loop condition (1) has to be fulfilled --- i.e. the
optimization task consist in finding the minimum of (2) under the
constraint (1)!

For $f_i(n)=f(n)$ independent of the bond index the minimum is
trivial: $n=0$, i.e.\ no bond is occupied. Only if the minima of local cost
functions vary from bind index to bind index in a non-trivial way one
can expect a non-trivial loop configuration. We assume a random
distribution of these minima (at values $b_i$) and restrict ourselves
to a quadratic form of $f_i(n)$ around these minima:
$f_i(n_i) = (n_i-b_i)^2$, which means that we study the 
loop configurations (i.e. occupied bond configurations that fulfill
(1)) that minimize 
\be
 H = \sum\limits_i (n_i - b_i)^2\;.
\label{hamiltonian}
\ee
The random variables $b_i$ are uniformly distributed $b_i \in
[-2\sigma , 2\sigma]$ at a fixed {\it disorder strength} $\sigma \in
[0,1]$.  Here, the probability $p$ to occupy a bond depends on the
disorder strength $\sigma$.  We refer to the LP model {\it with} an
optimization constraint of Hamiltonian (\ref{hamiltonian}) as the {\it
loop Hamiltonian} model.  For the minimization of a free energy
(optimization) we restrict to the calculation of the ground state
($T=0$) configuration ${\bf n}$, which is a minimum cost flow problem
that can be solved {\it exactly} in polynomial time with appropriate
algorithms\cite{HR02}.

\begin{figure*}
\includegraphics[width=17cm]{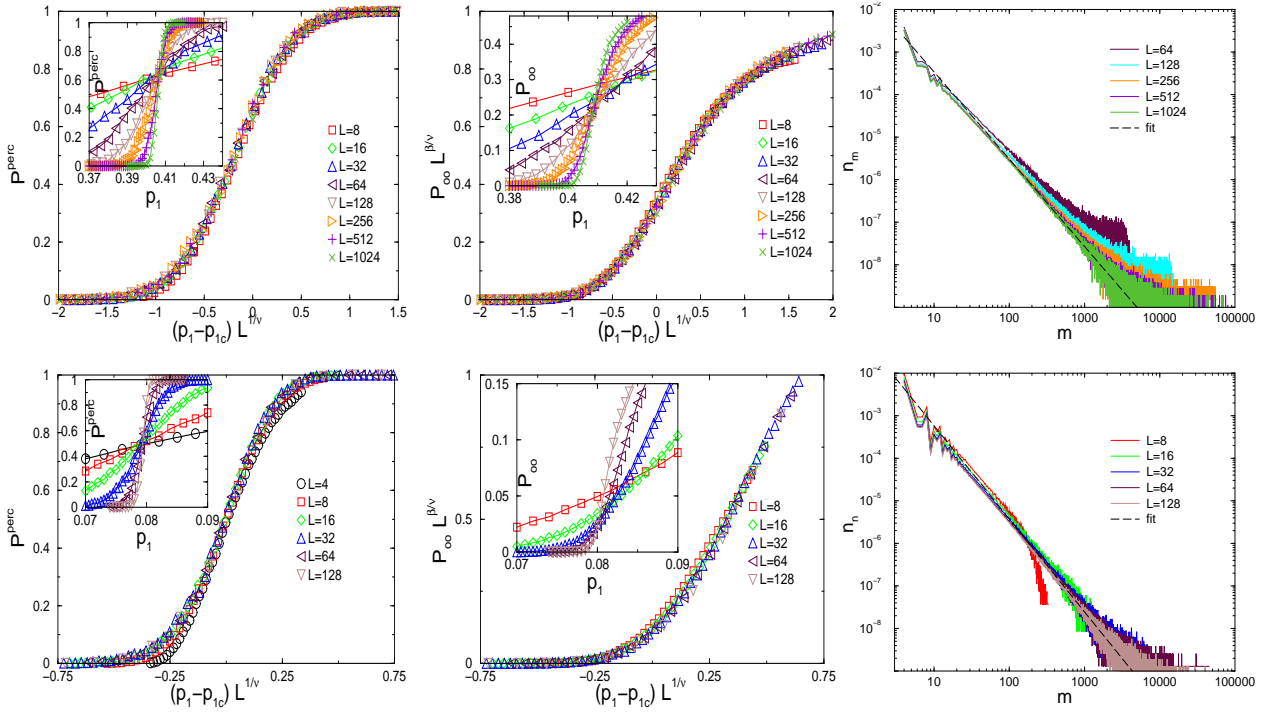}
\caption{
Finite-size scaling (FSS) for the LP model 
in $2d$ (top) and $3d$ (bottom):
Plot of the percolation probability $P^{perc}$ (left) and of the
probability $P_{\infty}$ for a bond belonging to a percolating loop
(middle).
The inset shows the raw data.
(Right) Plot of the average number $n_m$ of loops of mass $m$ per
lattice bond at $p_{1c}=0.407$ in $2d$ and at $p_{1c}=0.0793$ in $3d$,
respectively.
}
\label{2d1_Perc_Pinf}
\label{1_nm}
\end{figure*}

\section{Results}
\label{results}

We use a {\it depth-first search} algorithm known from combinatorial
optimization \cite{HR02} to identify the connected loops. The number
of realizations we used to get statistically averaged data varied from
500 for the largest system size to 20000 for the smallest system
size. In the following the error bars of our data in the figures are
smaller than the symbol size and are therefore omitted.

Fig. \ref{conf_3d3} and \ref{conf_hm} depict three typical loop
configurations of the LP and LH model around the critical threshold,
respectively, which clearly indicates a percolation phase transition
for both models.

\subsection{Loop percolation model (LP)}

First, we study the LP model and consider the probability $P^{perc}$
that a loop percolates the system.
Since we assume to have only one typical length scale, which diverges
at the critical point like $\xi \sim |p_1-p_{1c}|^{-\nu}$,
in a finite system $P^{perc}$ is expected to scale like
\be
P^{perc}(L)\sim \bar{P}[(p_1-p_{1c}) \; L^{1/\nu}].
\label{Perc_sc1}
\ee
Thus, $P^{perc}(L)$ is independent of $L$ at $p_{1c}$ and the data
curves should intersect for different system sizes $L$.
From our raw data in the inset of Fig.\ref{2d1_Perc_Pinf} (left) we
locate the critical point $p_{1c} = 0.4070 \pm 0.0005$ in $2d$ and
$p_{1c} = 0.0793 \pm 0.0004$ in $3d$, respectively.
We plot the scaling assumption (\ref{Perc_sc1}) in
Fig. \ref{2d1_Perc_Pinf} (left) and estimate the inverse correlation length
exponent $1/\nu = 0.75 \pm 0.03$ in $2d$ and $1/\nu = 1.143 \pm 0.090$
in $3d$ from the best data collapse at fixed $p_{1c}$.

To get a second critical exponent we consider the probability
$P_{\infty}$ that a bond belongs to the percolating loop, i.e. the
order parameter, which is expected to obey
\be
P_{\infty}(L) \sim L^{-\beta / \nu}  \bar{P}[(p_1-p_{1c}) \; L^{1/\nu}].
\label{Pinf_sc1}
\ee
Fig. \ref{2d1_Perc_Pinf} (middle) shows the raw data (inset) and the
plot of the scaling law (\ref{Pinf_sc1}) with $\beta/\nu = 0.104 \pm
0.020$ in $2d$ and $\beta/\nu = 0.49 \pm 0.02$ in $3d$ such as to achieve
the best data collapse.
From the $\nu$ above we determine $\beta$ shown in Table \ref{table}.

At the critical point $p_{c1}$ the average number $n_m$ of finite
loops of mass $m$ per lattice bond scales like
\be
n_m(L,p_1=p_{1c}) \sim m^{-\tau},
\ee
where $\tau$ is the Fisher exponent \cite{Sta85BH96,Fis67}.
Since we assume the usual scaling relations of conventional
percolation to be valid \cite{Sta85BH96}, we also expect a combination
of them to be valid, i.e. the hyperscaling relation
\be
\tau = \frac{2 - \beta/(d \nu)}{1 - \beta/(d \nu)},
\label{tau_sc}
\ee
where $d$ is the spatial dimension.
From the fit of the data of $n_m(L)$ at $p_{c1}$ in Fig. \ref{1_nm}
(right) we get $\tau$ depicted in Table \ref{table}.
This is consistent with the value from putting the above $\nu$
and $\beta$ into Eq. (\ref{tau_sc}), i.e. $\tau = 2.05 \pm 0.05$ in $2d$
and $\tau = 2.20 \pm 0.06$ in $3d$.

To determine the critical probability $p_c$ that a bond is occupied we
calculate
\be
p_c(L) = \sum_{m=4}^{m_L} m \; n_m(L;p_1=p_{1c}),
\label{p_c_eq}
\ee
where $m_L$ is the largest finite loop.
We plot $p_c(L)$ versus $1/L$ as depicted in Fig. \ref{pc_1L} and
extract $p_c$ from the limit $L \to \infty$.

In addition to the results presented so far, we found that the mean
number $N^{perc}$ of percolating loops per sample in the finite system
can be described by a smeared step function with an upper boundary
$N^{perc}=1$ as known from conventional percolation \cite{Sta85BH96}.
When we define the mass $m_i$ of an occupied bond $i$ to be $m_i=1$
even for $|n_i|>1$, the critical scaling behavior remains unchanged
and the critical probability becomes $p_c=0.565 \pm 0.005$
in $2d$ and $p_c = 0.266 \pm 0.005$ in $3d$, respectively.
We also studied the case, where the algorithm detects the loops
along oriented paths, and found the same results, as expected from what
we said above.

\begin{figure*}
\includegraphics[width=17cm]{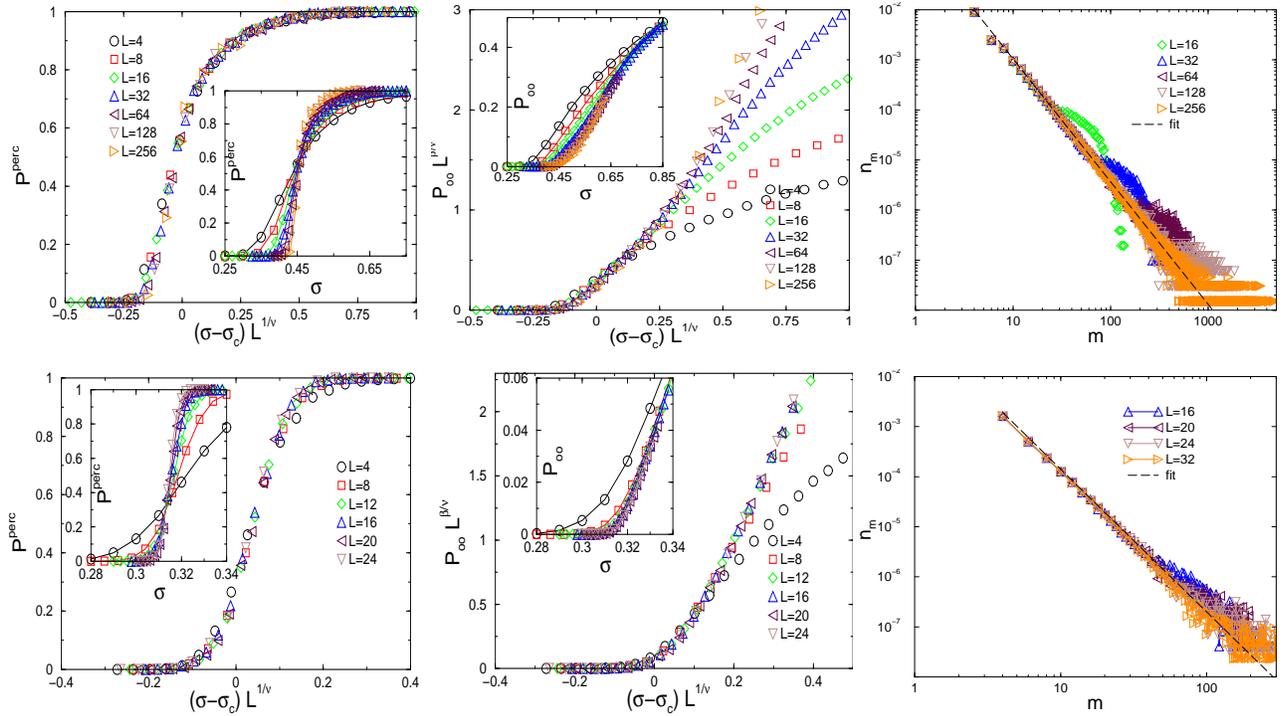} 
\caption{
FSS for the loop Hamiltonian (LH) model in $2d$ (top) and $3d$ (bottom). 
Plot of the percolation probability $P^{perc}$ (left) and of the
probability $P_{\infty}$ for a bond belonging to a percolating loop
(middle).
The inset shows the raw data.
(Right) Plot of the average number $n_m$ of loops of mass $m$ per
lattice bond at $\sigma_c=0.458$ in $2d$ and at $\sigma_c=0.3129$ in $3d$,
respectively.
}
\label{SOSPercPinf}
\label{hm_Perc_Pinf}
\label{hm_nm}
\end{figure*}

\subsection{Loop hamiltonian model (LH)}

For the loop Hamiltonian (\ref{hamiltonian}) we perform analogous data
analysis.  From the intersection of the $L$-dependent curves of
$P^{perc}(L)$ in the inset of Fig. \ref{hm_Perc_Pinf} (left) we locate
the critical disorder strength at $\sigma_c = 0.458 \pm 0.001$ in $2d$
and $\sigma_c = 0.3129 \pm 0.0005$ in $3d$, respectively.  From the
finite-size scaling behavior of the percolation probability $P^{perc}$
(similar to Eq. (\ref{Perc_sc1})) we get $1/\nu = 0.30 \pm 0.05$ in
$2d$ and $1/\nu = 0.95 \pm 0.05$ in $3d$.  The resulting exponent
$\nu$ are given in Table \ref{table}.  In $2d$ we find a value
$\nu=3.3 \pm 0.3$, which is rather large.

Fig. \ref{hm_Perc_Pinf} (middle) shows the plot of the raw data of
$P_{\infty}$ (inset) and its scaling law similar to
Eq. (\ref{Pinf_sc1})   
with $\beta/\nu = 0.55  \pm 0.05$ in $2d$ and $\beta/\nu = 1.30 \pm
0.05$ in $3d$.
From the $\nu$ above we determine $\beta$ in Table \ref{table}.

\begin{figure}
\includegraphics[width=\columnwidth]{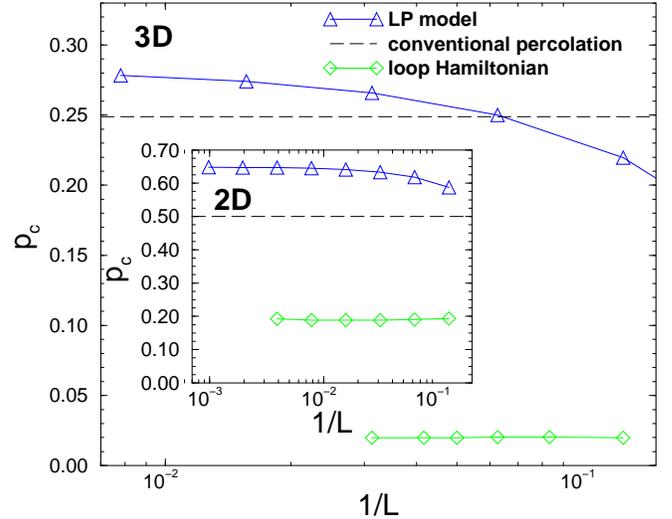}
\caption{
Plot of the critical probability $p_c(L)$ vs. 
inverse system size $1/L$  in $2d$ and in $3d$ for the
conventional percolation model, the LP model and the loop Hamiltonian
(LH) model.
}
\label{pc_1L}
\end{figure}
\begin{figure}
\bt{c}
\includegraphics[width=\columnwidth]{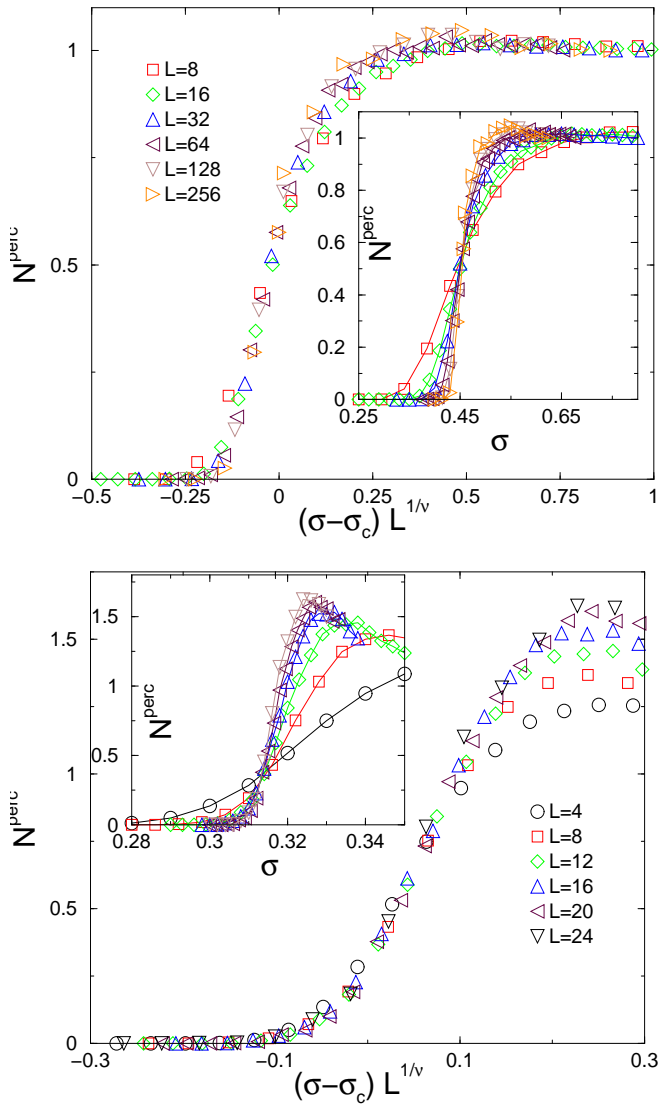}
\et
\caption{
Plot of the mean number $N^{perc}$ of percolating loops for the loop
Hamiltonian (LH) model in $2d$ (top) and $3d$ (bottom).
}
\label{hm_NPerc}
\end{figure}
\begin{figure}
\includegraphics[width=\columnwidth]{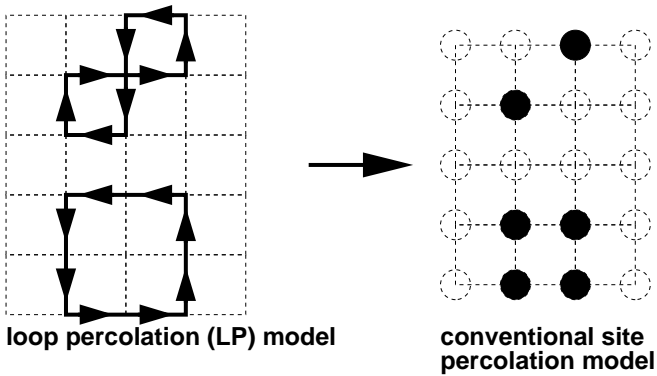}
\caption{\label{mapping}
  Schematic mapping of the loops in the LP model
  (left) onto  sites (filled circles) in the conventional site
  percolation model with next nearest neighbors (right; see text).
}
\end{figure}

In Fig. \ref{hm_nm} (right) we plot the loop distribution $n_m(L)$
vs. the mass $m$ at the critical point $\sigma_c$ and determine $\tau$
by power law fit, which gives the values shown in Table \ref{table}.
From $\nu$ and $\beta$ above we get via the hyperscaling relation
(\ref{tau_sc}) the Fisher exponent $\tau = 2.38 \pm 0.17$ in $2d$ and
$\tau = 2.76 \pm 0.26$ in $3d$, which are consistent with the values
from the power-law fit within the error bars.

In Fig. \ref{hm_NPerc} we plot the mean number $N^{perc}$ of
percolating loops per sample.
The different curves of $N^{perc}(L)$ intersect at the same critical
point $\sigma_c$ found for $P^{perc}$ above.
Similar to the scaling law of $N^{perc}_{conv.}(L)$ in conventional
percolation \cite{Sta85BH96} we expect $N^{perc}(L)$ to obey
\be
N^{perc}(L) \sim \bar{N}[(\sigma-\sigma_c) \; L^{1/\nu}],
\ee
and estimate the same $\nu$ as above from the best data collapse, see
Fig. \ref{hm_NPerc}.  

The mean number $N^{perc}$ of percolating loops per sample can become
larger than one slightly above the critical point $\sigma_c$ in
contrast to conventional percolation, where only one percolating
cluster exists for $p>p_c$. The appearance of several percolating
loops can possibly be related to the fact that the loop density in the
LH model at $\sigma_c$ is much smaller than in the LP model, as can be
seen from comparison of typical loop configurations
(c.f. Fig. \ref{conf_3d3} and \ref{conf_hm}). Moreover, the maximum of
$N^{perc}(L)$ seems to increase with increasing $L$.  From our data we
could not determine the behavior of the maximum of $N^{perc}(L)$ in
the thermodynamic limit $L \to \infty$, in particular whether it
converges to a constant or diverges.  We also checked a $L$-dependent
power law behavior of $N^{perc}(L)$ similar to Eq. (\ref{Pinf_sc1})
with a new critical exponent $x$ (instead of $\beta$) and found
$x/\nu=0.01 \pm 0.01$ in $2d$ and $x/\nu=0.15 \pm 0.02$ in $3d$,
i.e. $x$ close to zero.

Finally, we calculate the probability $p_c(L)$ that a bond is occupied
(analogous to Eq. \ref{p_c_eq}) for different $L$ at the critical point
$\sigma_c$ as shown in Fig. \ref{pc_1L} and extract $p_c$ as the
average value, see Table \ref{table}.

In all considered ground state configurations of the loop Hamiltonian
(\ref{hamiltonian}) we found a bond to be empty or singly occupied
only.  Due to this observation we also investigated - beside the study
presented this paper - a modified LP model, in which a plaquette is
allowed to be occupied with $p_1$ if and only if the amount of the
resulting flow ${\bf n}$ is $|n_i| \le 1$.  Here, the algorithm
checked each plaquette to be occupied or not in positional order.
Note that this occupation process depends on the algorithmic
order of occupying plaquettes in the system.
Again, we found the same critical exponents as known from conventional
percolation, but a different critical probability $p_{1c}$.

We also studied the $2d$ LH model with a different probability
distribution function $P(b_i)$, where $b_i$ is given by a sum of two
uniformly distributed random numbers out of $[0,2\sigma]$.  This
corresponds to the solid-on-solid (SOS) model on a disordered
substrate, which has been studied \cite{RB97,PR00} only at
$\sigma=1/2$ yet.  For this probability distribution function we get
the same critical exponents as found above, but with a different
critical point at $\sigma_c=0.395 \pm 0.005$, i.e. $p_c=0.34 \pm
0.02$. This implies that our study is relevant to describe a
disorder-driven flat-to-superrough phase transition, not studied in
literature yet.

Closely related to the SOS model is the $2d$ model of a random elastic
medium with contour loops, for which Zeng et al. found \cite{ZKNM98}
the geometrical exponents $\beta/\nu = d-d_f =0.54 \pm 0.01$ and $\tau
= 2.32 \pm 0.01$ at $\sigma=1/2$.  These exponents agree with the
critical $\beta/\nu = 0.55 \pm 0.05$ and $\tau = 2.38 \pm 0.17$ we
found here at $\sigma_c\approx 0.458$.

\begin{table*}
\bt{llccccc}
\hline

   & & conventional & conventional bond & LH & LP & LP with singly\\
   & & percolation \cite{Sta85BH96} & percolation \cite{here} & & &
   occupied bonds\\
\hline
\hline

$2d$ & $p_c$ & $0.5927460^a$ and $1/2^b$ & $0.5000\pm 0.0004$ & $0.189 \pm 0.005$     & $0.650
 \pm 0.005$ &  $0.570 \pm 0.005$\\
 & $p_{1c}$ & & & & $0.4070 \pm 0.0005$ & $0.5485 \pm 0.0005$\\
 & $\sigma_c$ & & & $0.458 \pm 0.001$ && \\
   & $P^{perc}(p_c)$ & & $0.70\pm0.02$& $0.59\pm0.02$ & $0.64\pm0.02$ & $0.67\pm0.02$ \\
   & $\nu$ & $4/3=1.\bar{3}$ & $1.33 \pm 0.05$& $3.33 \pm 0.30$ 
   & $1.33 \pm 0.05$   & $1.33\pm 0.04$  \\
   & $\beta$ & $5/36=0.13\bar{8}$ &  $0.139\pm 0.030$ & $1.80 \pm 0.35$
   & $0.138 \pm 0.027$  & $0.139\pm 0.007$ \\
   & $\tau$   &      && $2.45 \pm 0.05$  & $2.05 \pm 0.10$ &\\
\hline

$3d$ & $p_c$ & $0.31161^a$ and $0.248814^b$ & $0.2489\pm 0.0002$ & $0.0198\pm 0.0005$ & $0.282 \pm0.005$  &  $0.267 \pm 0.005$\\
 & $p_{1c}$ & & & & $0.0793 \pm 0.0004$ & $0.0992 \pm0.0005$\\
 & $\sigma_c$ & & & $0.3129 \pm 0.0005$ && \\
   & $P^{perc}(p_c)$ & & $0.63\pm0.02$ & $0.34\pm0.02$ & $0.53\pm0.02$    & $0.54\pm0.02$ \\
   & $\nu$ & 0.875 &  $0.875\pm 0.070$ & $1.05\pm0.05$ &$0.875\pm 0.070$ & $0.875\pm 0.070$\\
   & $\beta$ & 0.417 & $0.43\pm 0.04$ & $1.4 \pm0.1$  & $0.43\pm 0.04$   & $0.42\pm 0.04$\\
   & $\tau$   &        &  & $2.85 \pm 0.05$  & $2.19\pm0.05$  & \\
\hline
\et
\caption{
Comparison of the critical thresholds and critical exponents for
conventional bond percolation, the loop percolation (LP) model and the
loop Hamiltonian (LH) model.
$^a$ refers to conventional site percolation and $^b$ refers to
conventional bond percolation.
}
\label{table}
\end{table*}

\section{Summary}
\label{summary}

We studied two loop percolation models, numerically: 
in the {\it loop percolation} (LP) model the loop configuration resulted from
an uncorrelated unbiased random occupation of elementary directed plaquettes,
while in the {\it loop Hamiltonian} (LH) model the loop
configuration appeared according to the Boltzmann weight of a
particular microscopic model at $T=0$, i.e. from an optimization
constraint (\ref{divfree}) of a Hamiltonian (\ref{hamiltonian}). Our results are
summarized in Table \ref{table}.

We found that in $2d$ and $3d$ the LP model belongs to the
universality class of the conventional (bond or site) percolation
\cite{Sta85BH96}.  A plausible explanation for this observation is the
following: We map an occupied (empty) plaquette onto a occupied
(empty) site on an appropriate lattice.  Fig. \ref{mapping}
illustrates the mapping in $2d$: the two loops on a square lattice of
system size $L$, Fig. \ref{mapping} (left), are mapped onto two
clusters consisting of 2 and 4 occupied sites on a (dual) square
lattice of size $L-1$, Fig. \ref{mapping} (right).  The resulting
clusters of sites are clusters of finite extended objects like k-mers
in Ref. \cite{LP94}, which show the same universal behavior at the
percolation transition as conventional site percolation
\cite{Sta85BH96}. Also we found that the non-trivial orientation of
the loops in the LP model is irrelevant for the universality class.
We expect that this also holds for higher spatial dimensions $d$ with
the same critical dimension $d_c=6$ as for conventional percolation
\cite{Sta85BH96,Car97}.

For the loop Hamiltonian (LH) model Eq. (\ref{hamiltonian}) we found
evidence for an unconventional universality class of percolation
in $2d$ and $3d$, the exponents are listed in table I.  
In $2d$, we obtained a rather large correlation length exponent
$\nu = 3.3 \pm 0.3$ , which possibly indicates an infinite critical
exponent for $L \to \infty$ known from the Kosterlitz-Touless (KT)
phase transition \cite{KT}.
On the other hand, since our Hamiltonian (\ref{hamiltonian}) has no XY,
a Kosterlitz-Touless (KT) transition can be ruled out.
Indeed, from applying a KT-form of finite-size scaling to our data, we
could not find any acceptable data collapse.
We like to remark that loop percolation with a large correlation
exponent $\nu$ also appears in integer Quantum Hall systems, where
the loops represent equipotential lines \cite{CRSR01}.

It would be interesting to study the universal behavior of  the pure
(i.e. $\sigma=0$) LH model (\ref{hamiltonian}) for finite temperatures
$T$. In $3d$, such a thermal-driven loop percolation phase transition has been
studied \cite{NS98} for a different model.

\acknowledgements
We thank Jae Dong Noh for fruitful discussions and stimulating ideas.
This work has been supported financially by the Deutsche
Forschungsgemeinschaft (DFG).


\begin{references}
\bibitem{here}
  We reproduced the results of conventional bond percolation for
  comparison, where we used parameters (i.e. number of samples and
  system sizes) similar to the study of the LP model.
\end{references}
\end{document}